\documentclass[a4paper,twocolumn,aps,pra,showpacs,floatfix]{revtex4}

\usepackage{graphicx}
\usepackage{amsmath}
\usepackage{amstext}

\def\sx{\sigma_x}
\def\sy{\sigma_y}

\def\eig{\mathop{\mbox{eig}}\nolimits}

\begin{document}
\title{Universal covariant quantum cloning circuits for qubit entanglement manipulation}

\author{Levente Szab\' o}
\affiliation{Institute of Physics, University of P\'ecs, H-7624
  P\'ecs, Ifj\'us\'ag \'utja 6, Hungary}
\author{M\'aty\'as Koniorczyk}
\affiliation{Institute of Physics, University of P\'ecs, H-7624 P\'ecs, Ifj\'us\'ag \'utja 6, Hungary}
\affiliation{Research group for high performance computing, SP! Project Office, University of P\'ecs, H-7633 P\'ecs, Sz\'ant\'o Kov\'acs J\'anos u. 1/B.}
\author{Peter Adam and J\'ozsef Janszky}
\affiliation{Institute of Physics, University of P\'ecs, H-7624 P\'ecs, Ifj\'us\'ag \'utja 6, Hungary}
\affiliation{Research Institute for Solid State Physics and Optics, Hungarian Academy of Sciences, H-1525 Budapest, P.O. Box 49., Hungary}

\date{November 20, 2009.}
\begin{abstract}
  We consider the entanglement manipulation capabilities of the
  universal covariant quantum cloner or quantum processor circuit for
  quantum bits. We investigate its use for cloning a member of a
  bipartite or a genuine tripartite entangled state of quantum
  bits. We find that for bipartite pure entangled states a nontrivial
  behavior of concurrence appears, while for GHZ entangled states a
  possibility of the partial extraction of bipartite entanglement can
  be achieved.
\end{abstract}
\maketitle

\section{Introduction}

As it is prevalently known, quantum mechanics forbids the copying of
individual quantum systems, however, it is still possible to create
imperfect replicas of a quantum state with optimal
fidelity~\cite{PhysRevA.58.1827}. This protocol, originally introduced
by Bu\v zek and Hillery~\cite{PhysRevA.54.1844}, is called quantum
cloning, and it has been studied very extensively (see
Ref.~\cite{RevModPhys.77.1225} for a review).

In this paper we are interested in universal cloning transformations
for individual quantum bits. A quantum circuit was designed by Bu\v
zek et al. for this purpose~\cite{PhysRevA.56.3446}, which was later
generalized to arbitrary dimensional quantum
systems~\cite{ISI:000168589100041}. We shall call this circuit UCQC
(universal covariant quantum cloner) in what follows. It has a special
feature of being quantum controlled, that is, the fidelity ratio of
the two clones is controlled by the initial quantum state of two
ancillary quantum bits (one of which will carry the clone after the
process). This idea turned out to be related to the concept of
programmable quantum networks or quantum
processors~\cite{PhysRevLett.79.321}. These are fixed quantum networks
which are capable of performing operations on quantum systems in a way
that the operation itself is encoded into the initial quantum state of
ancillae. It was found that the very circuit for universal quantum
cloning is in fact a probabilistic universal quantum
processor~\cite{PhysRevA.65.022301}.

In this paper we consider UCQC-s as entanglement manipulation
devices. In the context of cloning, one may ask several questions. One
may consider the cloning of an entangled quantum state as a whole, in
order to obtain similar entangled pairs. For two qubits this has been
analyzed in detail by by several
authors~\cite{PhysRevA.71.042332,PhysRevA.69.040301,PhysRevA.72.042314}.
In particular, Bu\v zek at al.~\cite{PhysRevLett.81.5003} compare the
fidelity of cloning of an entangled pair by global and local
operations.

Another approach might be the broadcasting of entanglement, proposed
by Bu\v zek et al.~\cite{PhysRevLett.81.5003}. In this case two
parties share an entangled pair and use cloners locally to obtain two
partially entangled pairs. This protocol attracted a relevant
attention in the literature, too. Topics such as state-dependent
broadcasting~\cite{ISI:000238571200015}, broadcasting of multipartite
entangled states:
W-states~\cite{ISI:000238571200015,ISI:000265087300009},
GHZ-states~\cite{ISI:000241067100050}, and linear optical
realizations~\cite{PhysRevA.72.032331} were discussed in detail.
Our present study is motivated mainly by these works.

Consider an entangled pair. It is always interesting to ask what
happens to the entanglement if any of the members of the pair is
subjected to some quantum information processing protocol. In the case
of quantum teleportation, for instance, rather strikingly the
teleported qubit inherits the entanglement of the original qubit with
its pair. It is rather natural to ask what happens in the case of a
universal quantum cloner. The answer for qubit pairs is partly given
by Bandyopadhyay and Kar~\cite{PhysRevA.60.3296}. They show that if a
member (or both members) of a maximally entangled qubit pair is
subjected to an optimal universal quantum cloning operation, the
resulting state is a Werner state. It is likely, however, that a
cloning transformation is realized by some quantum circuit, which uses
ancillae for carrying out the operation. It is obviously interesting
how the entanglement between the different quantum bits of such a
scenario (including also ancillae) behaves.  In this paper we consider
the UCQC as a circuit, not only the cloning operation itself. We
calculate entanglement as measured by concurrence. It turns out that
the ancillae play a very specific role and the behavior of concurrence
shows a rather interesting pattern. The recent optical realization of
certain programmable quantum gate arrays~\cite{micuda:062311} also
contributes to the relevance of this question.

Another similar question might be the partial extraction of bipartite
entanglement from a GHZ-type threepartite resource. It is known that
if a three qubits are in a GHZ state~\cite{phystoday199308_22}, then a
measurement on either of the three qubits in the $|\pm\rangle$ basis
(eigenbasis of the $\sx$ Pauli-operator) projects the state of the
remaining two qubits into a maximally entangled state. We show that if
the given particle is cloned in advance, it is possible to create
bipartite entanglement by measuring the clone, while there still
remains some purely threepartite entangled resource in the state of
the three parties. This is indicated by the possibility of entangling
a different pair of qubits by a next measurement. The nature of the
entanglement in the multipartite system can be also analyzed with the
aid of the Coffman-Kundu-Wootters
inequalities~\cite{PhysRevA.61.052306}, which quantify the monogamy of
entanglement. We shall present such an analysis, too.

This paper is organized as follows: in Section~\ref{sec:eprclone} we
analyze the behavior of bipartite entanglement in the case when UCQC
is applied to clone a member of a maximally entangled pair. In
Section~\ref{sec:ghz} we consider the application of UCQC for the
partial extraction of bipartite entanglement from a
Greenberger-Horne-Zeilinger state. In Section~\ref{sec:concl} the
results are summarized and conclusions are drawn.

\section{Bipartite pure states}
\label{sec:eprclone}

\begin{figure}
  \centering
\includegraphics[width=0.45\textwidth]{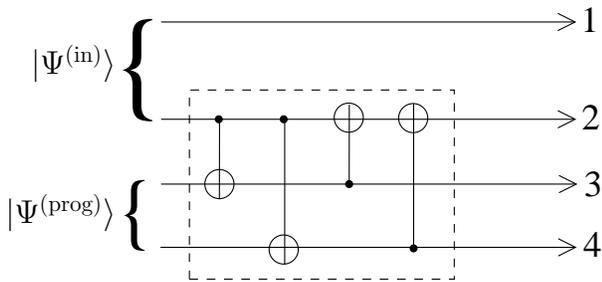}
\caption{The setup for bipartite entangled states. The dashed box
  contains the universal covariant cloning circuit, composed of four
  controlled-NOT gates.}
  \label{fig:setup}
\end{figure}

The considered setup is depicted in Fig.~\ref{fig:setup}.
The quantum circuit in the dashed box is the universal quantum
cloner~\cite{ISI:000168589100041}. Its first input port (which is
port 2 in our current notation) receives the state to be cloned,
while on the second two ports there impinges a so-called program
state:
\begin{equation}
  \label{eq:psiprog}
  |\Psi_{34}^{\text{(prog)}}\rangle= \mathcal{N} \left(
  \alpha \left( |0\rangle(|0\rangle  +|1\rangle)\right)
  +
  \beta  \left( |00\rangle +|11\rangle)\right)
  \right),
\end{equation}
where $\mathcal{N}=1/\sqrt{2(\alpha+\beta^2)}$ is a normalization constant
and $\beta=1-\alpha$. Were a single-qubit state $\varrho$
impingent on port 2, the output states would be:
\begin{eqnarray}
  \label{eq:cloneroutput}
  \varrho_2= \frac{\beta}{\alpha+\beta^2} \varrho 
  + \frac{\alpha^2}{2(\alpha+\beta^2)} \hat 1,
  \nonumber \\
  \varrho_3= \frac{\alpha}{\beta+\alpha^2} \varrho 
  + \frac{\beta^2}{2(\beta+\alpha^2)} \hat 1,
  \nonumber \\
  \varrho_4=  
  \frac{\alpha\beta}{\beta^2+\alpha} \varrho ^T +
  \frac{\alpha^2+\beta^2}{2(\alpha+\beta^2)} \hat 1
  .
\end{eqnarray}
The clones reside in ports 2 and 3, the original qubit and the first
ancilla, whereas in the port 4 there is an ancilla, the state of which
is a mixture of the state described by the mixture of the transpose of
the density operator of the original state and the identity
operator. The fidelity of the clones depends on the value of $\alpha$:
for $\alpha=0$ there is no cloning, whereas for $\alpha=1$ the state
of the original qubit is fully transferred to the clone, leaving the
original qubit in a completely mixed state. For other values of alpha
there are optimal clones generated. Note the symmetry of the formulae
in $\alpha$ and $\beta$.

Let us return to the description of the whole scenario in argument,
depicted in Fig.~\ref{fig:setup}. The qubits 1 and 2 carry the initial
bipartite input state. Qubit 2 is subject to cloning, while qubit 1,
the first part of the pair is not manipulated. We are interested in
the entanglement relations between the different pairs of qubits in
the resulting state.  As for the measure of bipartite entanglement for
qubits, we apply concurrence according to the Wootters formula: for a
quantum state $\varrho$ of two qubits,
\begin{eqnarray}
  \label{eq:concurrence}
  C(\varrho)=\max
  (0,\lambda_1-\lambda_2-\lambda_3-\lambda_4),\nonumber \\
  \lambda_i=\eig _i( \sqrt{ \sqrt{\varrho} \tilde \varrho \sqrt{\varrho}}),
\end{eqnarray}
where $\tilde \varrho=\sigma_y \otimes \sigma_y \varrho^\ast\sigma_y
\otimes \sigma_y$ is the Wootters-tilde, $\sy$ is the second
Pauli-operator, and $\varrho^\ast$ is the transpose of the density
matrix $\varrho$ in the product basis~\cite{PhysRevLett.80.2245}.

\begin{figure*}
  \centering
  \includegraphics{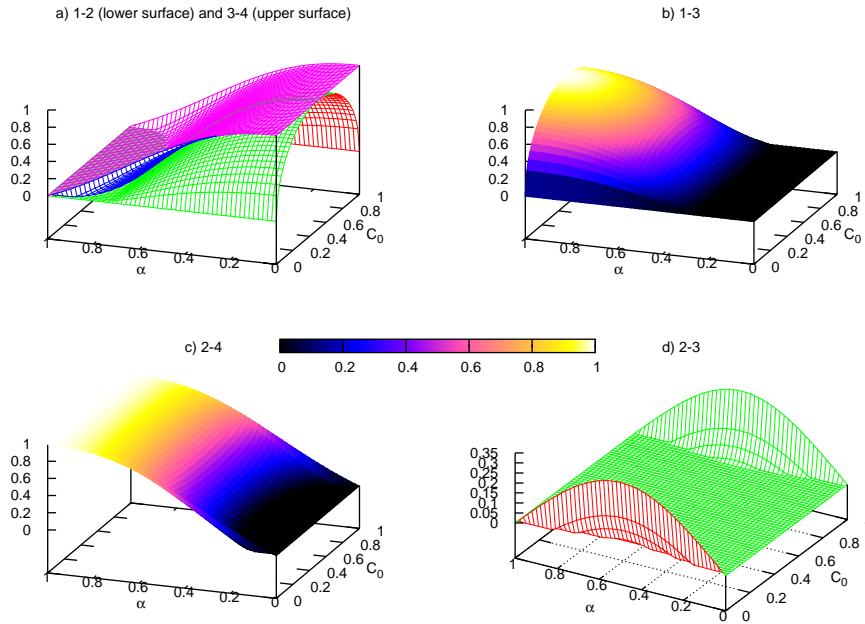}
  \caption{(color online) The entanglement between the various pairs
    of qubits at the output of the setup in Fig.~\ref{fig:setup}. The
    input state is either of the four in Eq.~\eqref{eq:instate2}, the
    same figure is obtained for each choice, though the states
    themselves differ. The ``program'' state is the one in
    Eq.\eqref{eq:psiprog}. The concurrence between qubits 1 and 4 is
    zero.}
  \label{fig:epr3d}
\end{figure*}

As an input state we consider a state in either of the following four
forms:
\begin{eqnarray}
  \label{eq:instate2}
  |\Phi_{12}^{\text{(in)}}\rangle= \sqrt{C_0} |00\rangle \pm \sqrt{C_1}
  |11\rangle, \nonumber \\
  \text{or} \nonumber \\
  |\Psi_{12}^{\text{(in)}}\rangle= \sqrt{C_0} |01\rangle \pm \sqrt{C_1}
  |10\rangle,\nonumber \\
\quad C_0+C_1=1. 
\end{eqnarray}
As for the nonzero concurrences between the various pairs of qubits, we
obtain the behavior in Fig.~\ref{fig:epr3d}, regardless of the choice
from the above states. The output states, however, depend on this
actual choice, we shall comment on this later.

In the figure one can observe that the entanglement between qubits 1
(the one not manipulated) and qubit 2 (the original qubit) behaves in
the similar way as that between qubit 1 and 3 (the one not manipulated
and the clone). For $\alpha=0$ (no cloning), qubits 1 and 2 are
entangled as they were originally, while for $\alpha=1$, complete
cloning, the entanglement is transferred to qubits 1 and 3, the clone
plays the role of the former original qubit completely. The surfaces
representing the concurrence for the pairs 1-2 and 1-3 are symmetric
in the cloning parameter $\alpha$, that is, they can be obtained from
each other by the $\alpha \to 1-\alpha$ substitution. The dependence
of these entanglements from $\alpha$ is monotonous but not continuous:
for small values there is a region where the entanglement is zero, and
it appears suddenly and non continuously. The dependence of these
concurrences on the initial entanglement in the state $C_0$ is
monotonous and continuous.

It is also interesting to observe that a similar non-symmetric
behavior appears in the concurrence of qubits 3-4 and 2-4. The program
state of Eq.~\eqref{eq:psiprog}, in which the qubits 3-4 are prepared
initially, is maximally entangled for $\alpha=0$, the case of no
cloning, and its entanglement decreases with the increase of the
cloning parameter $\alpha$. Accordingly, the entanglement of qubits
3-4 decreases with $\alpha$ also after the cloning operation, while
the complementary behavior (in the sense of $\alpha \to 1-\alpha$
substitution) appears between qubits 2 and 4 (the cloned part of the
input state and the ancilla of the cloner). Note that the entanglement
of qubits 3 and 4 is not equal to their entanglement \emph{before} the
cloning operation: the concurrence of the partially entangled program
state in Eq.~\eqref{eq:psiprog} is a monotonous and continuous function of
$\alpha$, and its values are not equal to the concurrences after the
cloning operation. Moreover, the concurrence of qubits 3 and 4 after
the cloning also depends slightly on that of the input state of qubits
1 and 2, in Eq.~\ref{eq:instate2}.

As for the remaining pairs, qubits 1 and 4 (the qubit not
manipulated and the ancilla) will not be entangled, while between
qubits 2 and 3 (the second input state and its clone), as a nontrivial
effect, there is a small amount of entanglement appearing only in the
case the input state (of qubits 1-2) is only slightly entangled.

\begin{figure}
  \centering
  \includegraphics[width=0.45\textwidth]{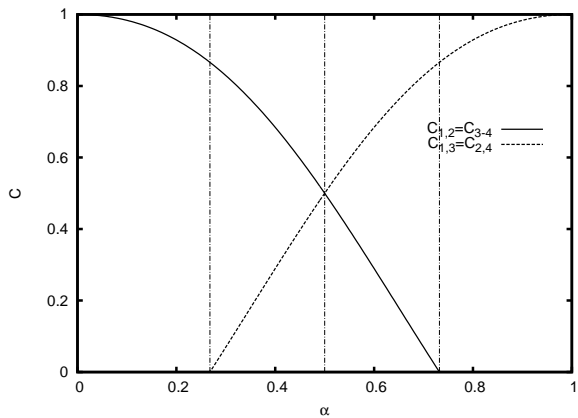}
  \caption{A slice of Figs.~\ref{fig:epr3d} a-c) for $C_0=1/2$, that is,
    for any of the maximally entangled Bell-states as
    input.}
  \label{fig:eprconc}
\end{figure}

A special case arises if the input state of qubits 1 and 2 is
maximally entangled. This is the case of $C_0=1/2$ in
Fig~\ref{fig:epr3d}. The concurrence between qubits 1-2 and 3-4 (two
originals, two program qubits of the cloner) is equal to each
other. The complementary pairs, qubits 1-3 (not manipulated-clone) and
2-4 (clone-ancilla) have also equal concurrences. The dependence of
these concurrences on $\alpha$ is depicted in Fig.~\ref{fig:eprconc}.
The behavior of these curves is due to the fact that the universal
cloning transformation produces Werner states. Indeed, if the input
state is in Eq.~\eqref{eq:instate2} is the maximally entangled
$|\Phi^{(+)}\rangle$ Bell-state, where the states of qubit-pairs 1-2,
1-3, 2-4, 3-4 are Werner-states of the form
\begin{equation}
  \label{eq:werner}
   \varrho^{(\text{Werner})}=\gamma
  |\Phi^{(+)}\rangle\langle\Phi^{(+)}| +
\frac{1-\gamma}{4} \hat 1,
\end{equation}
where $\hat 1$ stands for the identity operator of the two-qubit
space.  The value of the parameter $\gamma$ is
\begin{equation}
  \label{eq:gamma1}
  \gamma_{12}=\gamma_{34}=\frac{\alpha}{\alpha+\beta^2}
\end{equation}
for pairs 1-2 and 3-4, while it is 
\begin{equation}
  \label{eq:gamma2}
  \gamma_{13}=\gamma_{24}=\frac{\beta}{\alpha+\beta^2}.
\end{equation}
Note that the denominator on the right-hand-side of the above formulae
comes directly from the normalization constant of the program state in
Eq.~\eqref{eq:psiprog} (i.e. the scaling of parameters in
Eqs.~\eqref{eq:gamma1} and~\eqref{eq:gamma2} is merely a consequence
of our particular choice of parameters). In the case we choose a
different one from the states in Eq.~\eqref{eq:instate2}, we obtain
local unitary transforms of the Werner state in Eq.~\eqref{eq:gamma1}.
The message of the consideration for cloning an element of a maximally
entangled pair is not the fact that Werner states are obtained in
qubits 1-2 and 1-3, since it was known from the
literature~\cite{PhysRevA.60.3296}. What is nontrivial here that in
the UCQC circuit this behavior is repeated between the ancilla (qubit
4), and qubits 2 and 3, and this holds only in the case of the cloning
of a member of a maximally entangled state.  Finally let us note that
the behavior of qubits 2, 3, and 4 cannot depend on the properties of
qubit 1 since it is a remote system from the UCQC's point of view. It
is the reduced density operator of qubit 2 which can influence their
behavior. We have found that only for a maximally entangled pair,
a concurrence characterizing a nonlocal property is equal to another
concurrence which is a local property of the cloner.

\section{The GHZ state}
\label{sec:ghz}

In this section we consider the case in which a member of a
Greenberger-Horne-Zeilinger (GHZ) state is cloned. This tripartite
state, of the form
\begin{equation}
  \label{eq:GHZ}
  |\Psi^{(GHZ)}\rangle=\frac1{\sqrt{2}} (|000\rangle + |111\rangle)
\end{equation}
is known to be genuinely tripartite entangled. That is, all the
pairwise entanglements (as measured by concurrence) are zero, however,
all of the three qubits are in a maximally entangled state. When any
of the qubits is subject to a von Neumann measurement in the basis
\begin{equation}
  \label{eq:pmbasis}
  |\pm\rangle=\frac1{\sqrt{2}} (|0\rangle\pm |1\rangle),
\end{equation}
the other two qubits will be in either of the maximally entangled
Bell-states
\begin{equation}
  \label{eq:phipm}
  |\Phi^{\pm}\rangle = \frac1{\sqrt{2}} (|00\rangle \pm |11\rangle),
\end{equation}
depending on the measurement result. The probability of the measurement
results are equal. In this way the tripartite entangled resource in
the GHZ state can be converted into maximal bipartite entanglement.

\begin{figure}
  \centering
\includegraphics[width=0.45\textwidth]{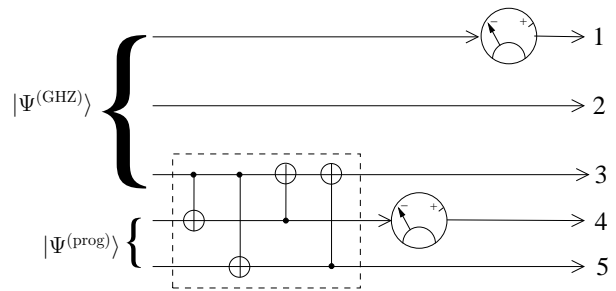}
\caption{The setup for the tripartite GHZ state. Qubits 1-3 hold the
  GHZ state initially. The dashed box contains the universal covariant
  cloning circuit, composed of four controlled-NOT gates. The
  ``meters'' measure in the $|\pm\rangle$ basis. First the clone
  (qubit 3), then a member of the original GHZ state (qubit 1) is
  measured. The horizontal position from the left to the right side
  thus represents the order of operations in time.}
  \label{fig:setupghz}
\end{figure}

The scenario we consider for GHZ states is depicted in
Fig.~\ref{fig:setupghz}.  Qubits 1-3 carry the input state which is a
GHZ state in Eq.~\eqref{eq:GHZ}. Qubit 3 enters the UCQC's first
port. The program ports of the UCQC are qubits 4 and 5, considered
again to be in the program state in Eq.~\eqref{eq:psiprog}.

\begin{figure}
  \centering
  \includegraphics[width=0.45\textwidth]{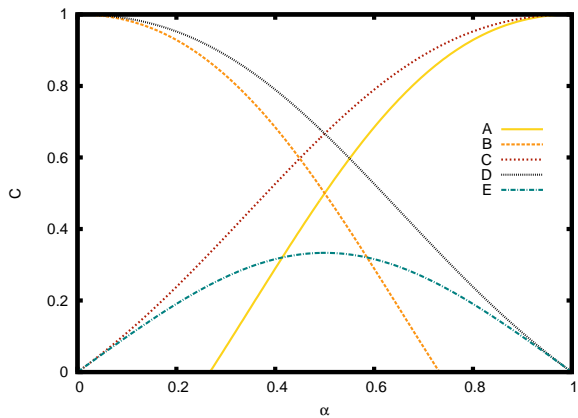}
  \caption{(color online) Pairwise concurrences in the GHZ-cloning scenario. A:
    qubits 3-5 after the cloner and also after each measurement, B:
    qubits 4-5 after the operation of the cloner, C: qubits 1-2 after
    the first measurement, D: qubits 2-3 after the second measurement,
    E: qubits 2-5 after the second measurement.}
  \label{fig:ghzconc}
\end{figure}

Directly after the operation of the cloner all pairwise concurrences
are zero, except for the one between qubits 3-5 and 4-5. Their value
is represented by curves ``A'' and ``B'' in
Fig.~\ref{fig:ghzconc}. This fact is easily explained by the following
reasoning. From the point of view of the UCQC circuit, qubits 1 and 2
are remote ones, thus they cannot influence the local properties of qubits
3, 4 and 5. All we ``see'' at the locus of the UCQC is that qubit 3 is
in a maximally mixed state, as it is a member of the maximally
entangled threepartite GHZ state of Eq~\eqref{eq:GHZ}. But the same
situation would arise if qubit 3 were maximally entangled in a bipartite
sense with one additional qubit, as we have considered in the previous
Section. Thus the behavior of concurrences between the pairs ancilla-original
and ancilla-clone are the very same as in the case of cloning a member
of a bipartite maximally entangled state (or a member of any kind of
multipartite entangled state which is itself in a completely mixed 
state for this reason): Werner states are obtained.

\paragraph{Projective measurement on the clone.}
Motivated by the relation of the projective measurements of the
members of a GHZ state on the $|\pm\rangle$ basis, one may now
consider a measurement of this kind on the clone, that is, on qubit 4.
This measurement will not alter the bipartite entanglement between
qubits 3-5, and that between 4-5 will disappear due to the
measurement. However, there will be an even larger entanglement
appearing between qubits 1 and 2, this is curve ``C'' in
~\ref{fig:ghzconc}. Both measurement outcomes will have equal
probability and also the entanglement behavior is the same for both
cases. In case of full cloning ($\alpha=1$), we obtain a pure EPR pair
as expected. (We remark here that if we were to measure on the
original qubit (qubit 3) instead of its clone, we would obtain the
counter propagating curve of the same shape, curve ``D'' in
Fig.~\ref{fig:ghzconc}, as one would expect. The role of the original
and the clone is symmetric. Entanglement of 4-5 will not alter, while
that of 3-5 will disappear in this case.) This is a partial conversion
of the resource available as genuine tripartite entanglement into
bipartite entanglement.

In order to further justify this statement let us consider a second
measurement, now on qubit 1. Again, the results will be uniformly
distributed and the entanglement itself will not depend on the
measurement result. The entanglement between qubits 3 and 5 will be
untouched, and that between qubits 1 and 2 will be destroyed by the
measurement of course. Meanwhile we obtain nonzero entanglement
between pairs 2-3 and 2-5, these are the curves ``D'' and ``E''
in Fig.~\ref{fig:ghzconc}, respectively. Indeed, if the extraction of
the tripartite entanglement was not full (i.e. $\alpha \neq 1$, one
can still obtain bipartite entanglement by measuring another qubit
this time. Curves ``C'' and ``D'', describing the entanglement between
1-2 after the first measurement, and 2-3 between the second,
respectively, are counter propagating, reflecting the interplay between
the two extractions. As a side effect, there is a small amount of
entanglement which appears between qubits 2-5 after the second
measurement, this is curve ``E'' in Fig.~\ref{fig:ghzconc}.

The use of the partial extraction of the entanglement is the
following. Consider that qubit 1 is at Alice, qubit 2 at Bob, while
the rest of the qubits is at Charlie. Initially they share a
tripartite GHZ resource. Charlie wants to enable Alice and Bob to use
a bipartite maximally entangled channel. He might perform the
projective measurement on the clone he has, however, in this case his
qubit 3 gets disentangled from the rest of the parties. However, if he
performs cloning and measures the clone, Alice and Bob still obtains a
partially entangled bipartite resource. However, Alice can decide that
instead of using a bipartite channel with Bob, she wants to create a
channel between Bob and  Charlie. All she has to do is to perform a
proper measurement on her qubit and communicate the result: Bob and
Charlie shall posses a partially entangled bipartite resource. This
would not be possible without the cloning. The same could be done of
course by Bob, to enable the bipartite resource between Alice and
Charlie.

In order to obtain a deeper insight into the behavior of bipartite
entanglement in this multipartite system, it is worth examining the
Coffman-Kundu-Wootters inequalities: for a system of quantum bits in a
pure state it always holds that
\begin{equation}
  \label{eq:ckw}
  \tau_k \geq \sum_{l\neq k} C_{k,l}^2,
\end{equation}
where $\tau_k=4\det \varrho (k)$ is the one-tangle, the linear entropy
of the density operator of the $k$-th qubit. If these inequalities
are saturated, the bipartite entanglement is maximal.

\begin{figure}
  \centering
  \includegraphics[width=0.45\textwidth]{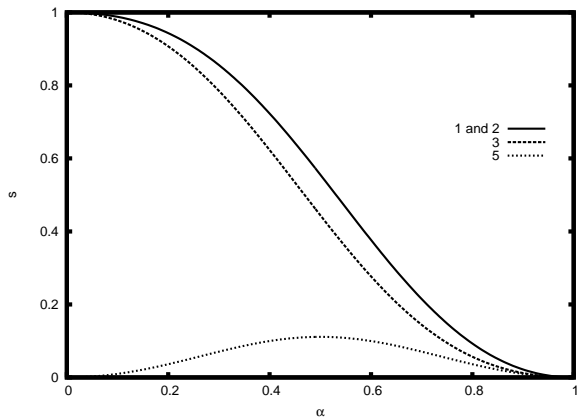}
  \caption{The quantity in Eq.~\eqref{eq:ckwsat}, which is zero if the
  CKW inequalities are saturated, for each qubit after the first
  measurement in our GHZ-cloning scenario. For qubit 4 the quantity is
zero.}
  \label{fig:ghzckw}
\end{figure}

To quantify the saturation we evaluate
\begin{equation}
  \label{eq:ckwsat}
  s=\tau_k-\sum_{l\neq k} C_{k,l}^2,
\end{equation}
which is zero if the inequalities are saturated.
After the first measurement we obtain nonzero values except for
the fourth qubit (apart from the case of $\alpha=1$. The behavior is
depicted in Fig.~\ref{fig:ghzckw}. The fact that the CKW inequalities
are not saturated also suggests the presence of additional
multipartite entanglement in the system. After the second measurement,
on the other hand, we find that all the CKW inequalities are saturated:
the system is in a sense maximally bipartite entangled.

\section{Conclusion}
\label{sec:concl}

We have shown that when using a universal covariant quantum cloning
circuit to clone a member of an entangled pair of qubits, a very
specific behavior of the entanglement of the qubits appears. The main
feature is that behavior of the entanglement between the not cloned
part of the pair and the cloned one is repeated in the entanglement of
certain ancillae, and so is that of the not cloned qubit and the
clone, provided that the original qubit pair was maximally entangled
initially. We have described the behavior of the entanglement in
detail.

We have also investigated the cloning of an element of the GHZ
state. It appears that the universal quantum cloning circuit
facilitates the partial extraction of bipartite entangled resources
from a genuine tripartite entangled resource. We provided a detailed
analysis of the entanglement behavior, including the relation to
Coffman-Kundu-Wootters inequalities.

In conclusion, the universal quantum cloning circuit (or quantum
processor) for qubits is found to be useful as an entanglement
manipulator as well. It can perform entanglement manipulations which
are potentially applicable in quantum information processing.

\acknowledgements

We acknowledge the support of the Hungarian Scientific Research Fund
(OTKA) under the contract No. T049234. M.K. acknowledges the support
of the project T\'AMOP 4.2.2. ``SP! IKT - Science, Please! Innovative
Research Team''.


\end{document}